\documentclass[10pt, conference]{IEEEtran}
\usepackage{amsmath}
\usepackage{amssymb}
\usepackage{array}
\usepackage{graphicx}
\usepackage{mathtools}
\usepackage{booktabs}
\usepackage{enumerate}
\usepackage{url}

\newcommand{\fsf}[1]{\textsf{\small{#1}}}

\DeclareMathAlphabet{\mathsl}{OT1}{ptm}{m}{sl}

\newcommand{\etal}{{et al.}}

\begin{document}

\title{True Peer Review}

\author{
 \IEEEauthorblockN{Amit K.~Chopra}
 \IEEEauthorblockA{
% Department of Information Engineering and Computer Science (DISI)\\
 University of Trento, Italy\\
%  Via Sommarive 14, Povo, Italy\\
 \fsf{chopra@disi.unitn.it}}}

\maketitle

\begin{abstract}

  In computer science, conferences and journals conduct peer review in
  order to decide what to publish.  Many have pointed out the inherent
  weaknesses in peer review, including those of bias, quality, and
  accountability.  Many have suggested and adopted refinements of peer
  review, for instance, double blind peer review with author
  rebuttals.

  In this essay, I argue that peer review as currently practiced
  conflates the sensible idea of getting comments on a paper with the
  irrevocably-flawed one that we either accept or reject the paper,
  which I term \emph{gatekeeping}.  If we look at the two separately,
  then it is clear that the ills associated with current peer review
  systems are not due to the practice of getting comments, but due to
  the practice of gatekeeping.

  \emph{True peer review} constitutes my proposal for replacing
  existing peer review systems.  It embraces the idea of open debate
  on the merits of a paper; however, it rejects unequivocally the
  exercise of gatekeeping.  True peer review offers all the benefits
  of current peer review systems but has none of its weaknesses.  True
  peer review will lead to a truly engaged community of researchers
  and therefore better science.

\end{abstract}

\section{The Debate on Peer Review}
\label{sec:intro}

``You just have to resubmit and hope to get assigned the \emph{right}
set of reviewers,'' advised an experienced mentor with whom I was
discussing ways of improving a recently rejected paper of mine.  In
other words, \emph{keep trying until you get lucky}.  Whereas the
advice was well-meaning, it betrayed a lack of trust in the peer
review process.  Others have echoed a similar sentiment.
Naughton~\cite{naughton:peer-review:2010} in a recent well-publicized
keynote mentions the large role a lucky assignment of reviewers plays
in getting a paper accepted.  Anderson~\cite{anderson:reviewing:2009}
backs up this claim with statistical evidence from computer
systems-related venues.

Many have noted the problems with traditional peer review.  Casati
{\etal}~\cite{casati:publish-perish:2007} criticize the current
publication model for entangling the separate concerns of
dissemination, evaluation, and retrieval.  In his keynote, Naughton
noted the problems resulting from the combination of the pressure to
publish, low acceptance rates, and poorly-trained reviewers, including
that of undue negativity in reviews .  More commonly noted are the
problems of bias and accountability and that most speculative,
potentially interesting research tends to get rejected in favor of
incremental work~\cite{fuggetta:universities:2012}.  A survey of
researchers undertaken on behalf of the Publishing Research Consortium
(PRC) contains extensive pointers to the ongoing debate on the
efficacy of peer review~\cite{ware:peer-review:2008}.

Researchers in computer science recognize some of the limitations of
peer review, and they are changing their systems to mitigate them.
For example, the AAMAS (Autonomous Agents and Multiagent Systems)
series of conferences implement double blind paper reviewing with
author rebuttals.  Further, senior program committee members and the
program chairs monitor the quality of the reviews.  Some conference
series such as ICSE (Software Engineering) and RE (Requirements
Engineering) have separate tracks for vision papers and new ideas and
emerging results.  The VLDB Foundation no longer publishes conference
proceedings: all papers accepted to the foundation's journal are
presented at the next VLDB conference.  Some have adopted more open
systems of peer review in order to counter the problem of
accountability, for example, the now defunct Electronic Transactions
on Artificial Intelligence.  One of Naughton's proposal for
improvement is not conducting peer review at all and accepting
everything.  Besides the engineering of peer review systems,
researchers have also attempted to educate potential reviewers and
writers on their respective tasks
\cite{smith:referee:1990,shaw:writing-papers:2003}.  Reviews forms at
most conferences and journals are increasingly detailed, ostensibly to
make sure that reviewers not overlook any major quality of the paper.

While these are all well-intentioned efforts, they miss the point: we
\emph{must} get out of the accepting-rejecting business altogether and
instead embrace the true spirit of scientific engagement, which I term
\emph{true peer review}.  The rest of this essay is an elaboration of
this point.

\section{Gatekeeping}
\label{sec:gatekeeping}

In computer science, conferences and journals conduct peer review in
order to decide what to publish.  Conferences and journals are, in
effect, institutions that perform the function of \emph{gatekeeping}:
the intent is to let in only good work.  What passes the gate is
\emph{published}.  Lest we get too hung up on adjectives, you can
replace ``good'' with your favorite adjective, e.g., ``interesting'',
``original'', ``solid'', and so on.

Implicit in gatekeeping is the notion that what is published is
\emph{authoritative}: worth knowing, worth citing, and worth building
upon.  What is not published is not worth knowing. (No conference or
journal that I know even publishes a list of rejected papers.) Not
published means low quality.  Conversely, published means high
quality.  Without these two axioms, gatekeeping would lose much of its
legitimacy.  The axioms may sound extreme but when I look around I see
most people and institutions behaving as if they were true.  Consider
for instance that one is neither likely to get tenure nor any research
funding without having published substantially.  Or consider, for
instance, that authors are not likely to cite anything except
published work.  Recently, I was criticized for citing workshop
papers, presumably because they are not as rigorously peer-reviewed.
Consider, for instance, that the recipe for success that is most
freely dispensed to junior researchers and faculty these days is not
``explore this theme deeper; it will lead to good results''; it is
``publish a lot in the top venues''.  Others before me have put it
more succinctly: \emph{publish or perish}.

This would all be well and good if gatekeeping were working.  There
are two problems with gatekeeping.  Whereas one is conceptual and
therefore more fundamental, the other concerns the way gatekeeping is
currently practiced.

\subsection{The Problem of Demarcation}
\label{subsec:demarcation}

Demarcating the good science from the bad is an enormously difficult
task.  In fact, if the philosophy of science has shown us anything, it
is that such value judgments are bound to be subjective: the review
depends on the reviewer \cite{kuhn:objectivity:1977}.  Each reviewer's
intellectual biases will inform his or her reviews.

We are all intellectually biased and our biases run so deep that we
may not even recognize them as such.  We all have our own
inspirations, our own beliefs, our own inclinations, and our own
favorite theories.  We all have different research backgrounds, with
some of us having worked in competing research paradigms.  We all
apply subtly different evaluation criteria by which we judge research.
We potentially favor different styles of exposition.  Our emotional
attitudes are also different; for instance, some of us may be more
forgiving of errors than others.  When one reviews a paper, he or she
brings all this to bear upon the review, but for the most part only
tacitly.  And yet we are inclined to claim objectivity!

We already know how deeply subjective peer reviewing is.  We know this
because different reviewers give different ratings for the same paper.
In fact, often enough, the reviews are blatantly conflicting.  Of
course, even with conflicting reviews, gatekeeping means that a
decision must be made.  So reviewers are encouraged by editors to
resolve their differences so that when the final decision goes out to
the authors, it would appear to have had unanimous support.  When
reviewers stick their ground, additional reviews may be solicited.
Then based on the reviews and the discussion, the program board (it
could be just the editor or program chair) somehow makes a judgment
call.  All this simply goes to highlight the subjectivity of peer
review.  It also goes to show how hard conferences and journals try to
create the illusion of objectivity where none exists.

Forget conflicting reviews.  Consider the case of unanimity. Do three
favorable reviews mean the work is objectively good and three
unfavorable ones that it is objectively bad?  If three other people
were to review the work, couldn't a ``good'' verdict turn ``bad'' and
vice versa?  If the whole world were to vote 'bad', it would likely
have social and psychological consequences for the authors, but that
still wouldn't make his or her paper bad.

That we are all intellectually biased is not a bad thing; it is simply
the way we are.  Kuhn, in fact, paints our intellectual biases in a
relatively positive light by explaining their value for problem
solving within a research paradigm~\cite{kuhn:revolutions:1962}.  Our
subjectivity is to be celebrated, not bludgeoned to death by having us
apply supposedly ``objective'' criteria to judge the merits of
research.

The simple point is that that while informed subjective viewpoints may
be valuable, they cannot serve as the means for objectively separating
the good from the bad.  Gatekeepers have taken upon themselves an
impossible task.  In fact, if we accept our subjectivity, then
gatekeeping actually gets in the way of progress.  There is simply no
need to put works published in conferences and journals on a pedestal
at the expense of others.

\subsection{The Problem of Accountability}
\label{subsec:accountability}

Peer review processes inspire little confidence in the authors because
no one on the reviewing side is accountable to the authors---not the
reviewers, not the program boards and program chairs at conferences,
and not the editors at journals.  The reviewers are anonymous and the
authors are not privy to any of the discussions on the reviewing side.
From the time that the author submits a paper to the time a decision
is made, he or she is completely out of the loop.

The authors may get a chance to respond to the reviewers but that
seldom has any effect.  With journals, there can be more of a back and
forth, but it is still quite limited.  The authors may complain to the
program chair or editor but that too rarely has any effect except for
a courteous reply saying the he or she must follow the recommendations
made by the reviewers.  

Why is accountability important? Some people believe that reviews by
and large are of good quality, but my experience as author, reviewer,
and program committee member has been to the contrary.  Many reviewers
simply repeat the authors claims followed by what would seem like an
arbitrary decision; many give all kinds of flimsy reasons that, if not
thoroughly unscientific in attitude, have nothing at all to do with
the contents of the paper.  Many reviewers simply follow a lexical
pattern-matching algorithm when doing a review (conferences and
journals, as mentioned before in Section~\ref{sec:intro}, provide
templates for writing reviews).  For example, there are those who are
so obsessed with experimental validation that I believe that they
would have told Dijkstra that his solution to the Dining Philosophers
Problem was wholly impractical because philosophers are known to be an
unruly bunch in general.

Further, interpersonal relationships and other psychological factors
likely play a big role in gatekeeping.  Aren't reviewers stung by
direct criticism and pleased by praise of their own work?  How common
is it that reviewers will recommend rejection of papers that take
positions contrary to their own and accept those that praise, extend,
or complement their own?  How commonly does the reputation of the
authors bear on the decision?  Don't many reviewers write their
reviews in a hurry?  Don't many (including people such as program
chairs and journal editors) not to want to revisit their original
reviews because of the extra work involved?  People want to be on
program committees and editorial boards but do they want to write
reviews?  Coupled with the fact that conferences want to have low
acceptance rates, doesn't the fact that many of the reviewers are also
authors produce a serious conflict of interest?  Many conferences and
journals say their review processes are rigorous and fair, but they
mean these things only in a very narrow bureaucratic sense, that is,
the steps of the peer review process were followed.

Good conferences and journals in computer science have acceptance
rates ranging from 15-25\%.  If we find a vast majority of the papers
unacceptable, why do we think that the vast majority of reviews would
be acceptable?  In my experience so far, editors and program chairs
are prone to turn a blind eye to author complaints about unfair
reviews; they justify doing that by saying the decision was arrived at
following due process.  How many times are reviews overturned?

\section{True Peer Review}
\label{sec:tpt}

What people currently understand as peer review conflates two things:
getting comments on a paper and gatekeeping, that is, deciding whether
it should be accepted or rejected.  The first is valuable and can
potentially lead to improvement in the paper.  \emph{The latter is
  just a case of getting our priorities wrong}.  Right now, we think
that we must build an authoritative source of knowledge, therefore we
must do gatekeeping.  Peer review is the means gatekeepers use to
justify their ends.  The first step towards solving the problems of
what is currently understood as peer review is to recognize
gatekeeping and peer review as two different activities.  For clarity,
I term the activity of peer review untangled from gatekeeping as
\emph{true peer review}.

True peer review is an exchange of informed opinions on a paper.  It
happens in a community of scientists.  True peer review is an
argumentative setting that also actively involves the author.  It
recognizes that comments, discussions, and arguments can potentially
lead to improvements in a paper.  It may be formal or informal.  We
often engage in true peer review over email when we send drafts and
papers to others for comments.  We participate in peer review when we
ask or get questions at research seminars.  Can anyone claim that the
reviews obtained from unaccountable anonymous randomly assigned people
are better than the questions and comments you receive from those
whose opinions you value and sought?  More formal true peer review
systems will involve third-party repositories of papers and
discussions about them.

Argumentation is the mechanism of true peer review.  It is likely that
no consensus would be achieved even after vigorous argumentation but
this is not a problem because even the exploration of different points
of view in an activity valuable in itself.  For a researcher, the
record of arguments could turn out to be just as rich a source of
information as the paper being argued about.  This record is as
valuable a scientific document as the paper it comments on.  It is a
pity that reviews from the current peer review system are for all
practical purposes \emph{forgotten}.

To engage in true peer review is up to individuals, whether in the
role of a reviewer or author.  Nobody can be forced to engage in it.
For example, an author may choose to not solicit any reviews; and
potential reviewers may turn down requests from the author.  However,
that hardly discredits true peer review.  If people tend not to write
reviews voluntarily, there is no reason to think they would do a good
job of it if forced.  If people don't solicit reviews on their work,
that doesn't stop anyone from either ignoring it or remarking upon it
in their own work.

Naturally, in any argumentative setting some participants may wield
more influence than others, but that is no different than current peer
review systems.  Unlike current peer review though, true peer review
is an end in itself, not the means to the irrevocably-flawed notion of
gatekeeping.

\section{False Arguments}
\label{sec:fa-gk}

Below, I go through a list of arguments that ostensibly make the case
for gatekeeping; however, I show them all to be false.

\emph{Reviewers at conferences and journals are more knowledgeable
  than the authors}.  If we could make objective claims about who is
more knowledgeable, we would be able to settle conflicts among
reviewers by that criterion.  We wouldn't even need three
reviewers---just one reviewer more knowledgeable than the author would
suffice.  The simple fact is we can't settle any difference of opinion
by resorting to claims of knowledgeability.  Further, consider that in
practice, papers are often reviewed by junior researchers, including
doctoral students (not a bad thing in itself, but it does undermine
the claim to authority).

\emph{Without gatekeeping, there would be no authoritative source of
  knowledge. Where would one even begin to look for information?} If
by `authoritative', one means worth-knowing, I already substantively
dismissed that argument in Section~\ref{sec:gatekeeping}.  Gatekeeping
produces results informed by personal biases, both intellectual and
political. In fact gatekeeping makes things worse: people will
potentially restrict themselves to a narrow body of published work.

One can turn this question around and ask if one must consider
everything that anyone bothers to write on a topic.  How can one
possibly cope? The simple answer is yes: ideally, one must do that
anyway.  Should a researcher not consider an unpublished report simply
because it was unpublished? Should he or she not consider it because
it was written by some hitherto unknown person? Should he or she not
consider it because it is only four pages long instead of ten?  It is
the \emph{ethical responsibility} of a researcher to consider
everything regardless of whether published or not or who the authors
are or what the format is.  In fact, it is researchers who attach a
high value to gatekeeping who are unlikely to meet this ethical
responsibility.

Even now we find a lot of information through our social networks,
which includes advisers, colleagues, collaborators, students, and so
on, and through Web search.  We often ask experts for references.  A
novel idea of dissemination that came out of the LiquidPub project
(\url{http://liquidpub.org/}) was that a researcher could publish his
or own journal.  The journal would contain papers written by others
that the researcher thought worth perusing, perhaps with his or her
own comments.  Journals published by experts would probably be more
visible than those published by relatively unknown researchers (a
precedent of this idea can be found in Dijkstra's unpublished
manuscripts).  I am confident that without gatekeeping, new efficient
ways of finding and keeping track of information will emerge.

Let me ask those who are frightened by what seems to them an immense
overload of information in a true peer review world: do they read
every relevant paper published in conferences and journals? If they
don't anyway, why bother to raise it as an argument against true peer
review?

\emph{Getting your paper published or not published is a choice.  One
  is free to not submit papers anymore to peer-reviewed venues.}  As
pointed out in Section~\ref{sec:gatekeeping}, given the extraordinary
importance accorded to published work, one does not really have a
choice.  True freedom is not about having choice, but about the
freedom free from pressure to exercise this choice.

\emph{The process of gatekeeping has produced many influential
  publications.}  There is no evidence for the claim.  I could argue
that the publications were influential only because they were
published or that they would have been influential even without
gatekeeping.

\emph{Given that people are biased anyway, we can't do much better
  than gatekeeping.}  There is no evidence for this claim.  A
counter-argument is that gatekeeping institutionalizes personal
biases.  In other words, instead of a person saying that ``I like (do
not like) this paper'', it is the institution (recall that by
'institutions', we mean conference and journals) that says ``We accept
(reject) this paper''.

Let personal biases be personal; a person can choose to make his or
her biases known in comments and reviews. But there is no reason to
make unaccountable anonymous reviewers' personal biases institutional.
Even if the reviewers' identity were made public, recall from
Section~\ref{subsec:demarcation} that reviewers are not authoritative
sources of knowledge.

\emph{But how can we judge the performance and merit of researchers
  without turning to publications? How can we make hiring and
  promotion decisions?}  I don't have a definite answer but I don't
think that the lack of gatekeeping makes these tasks any more
challenging.  For example, let's consider the task of hiring someone.
How would we do that?  By reading a few samples of what he or she has
written, by paying attention to the presentation of his or her work,
by listening to his or her vision of the future, by asking questions,
by interacting, by probing the depth and breadth of his or her
understanding, by judging his or her passion, by judging how well he
or she articulates his or her thoughts.  One cannot evaluate a
candidate on the basis of the broken system that is gatekeeping.  I
think hiring decisions already consider most of the above factors, but
publications are likely given a weight higher than any other factor
(hence the mantra \emph{publish or perish}), which I think is
misguided.

Consider that the current criteria for hiring have emerged because of
the importance we give to publications.  If we had no gatekeeping,
likely some other set of criteria would emerge.  One may argue that
looking at the publication record simplifies a difficult decision, but
that is an optimization given that our overriding \emph{value} is to
spend as little time and effort as possible on these decisions.  If
that value persists, the criteria that would emerge in the absence of
gatekeeping would also be in keeping with that value.

\emph{Given the limited time and resources at conferences, how do we
  decide which papers to present and which to not?} This is a separate
logistical problem similar to the one about hiring and promotions.  We
may have to rethink how we do conferences.  Perhaps there should only
be poster sessions at conferences.  Let researchers work on attracting
audiences to their posters; let them go and actually \emph{talk} to
others about their work rather than just do a 20-30 minute
presentation that few in the audience are interested in.  Solutions to
logistical problems will emerge.

\emph{Publications are incentives for researchers to produce high
  quality work.}  That seems like an absurd claim.  The best
researchers are driven by passion and the inklings of a superior way
of doing things.  They want to disentangle the threads, connect the
dots, fill the gaps, and turn things inside out.  It takes gatekeeping
(and systems of evaluation based upon it) to make a good researcher
produce bad work.

\emph{Personal biases, both intellectual and political, will not be
  eliminated in true peer review.}  True.  However, the bias is no
longer institutional.  There is no censorship.

\emph{True peer review cannot guarantee that every paper receives
  comments or reviews.}  True.  But balance that against the fact that
not all reviews one gets in the traditional system are useful.  True
peer review encourages you to seek comments from people who \emph{you}
think could provide useful comments.  True peer review encourages
people to engage with each other in meaningful discussions.  When
researchers discuss the relevant literature in their own report, they
are engaging in a limited form of true peer review (although,
unfortunately, the discussion of the literature is often a mere
formality in practice because of the political dimension of
gatekeeping).

As mentioned before, one could set up online repositories of papers
where people could comment and carry out debate on the merits of
papers therein.  Additionally, one could set up incentives to
encourage people to review the papers published there.  Comments could
be read and rated by others.  The best comments would filter to the
top and provide the commentator visibility.  The point is \emph{we can
  engineer systems to support true peer review}.

\section{Pragmatic Benefits of True Peer Review}

\begin{itemize}
\item True peer review saves time.  One does not have to make the
  changes one feels unnecessary merely to satisfy reviewers.  Authors
  can do the changes they feel necessary and move on to the next
  thing.  One will have more time and energy to pursue his or her own
  passions instead of being caught up in the revise and resubmit until
  accepted cycle.
\item Bid adieu to \emph{publish or perish}. Since there are no
  publications, the publication count becomes meaningless.  What
  becomes important is the author's message, both in its breadth and
  depth.  For example, in true peer review, one could deposit ten
  reports with minor changes among them.  However, he or she would
  have only one message to convey.  This also means that researchers
  will no longer have to recycle the same idea into more archival
  versions of the paper.  There simply will be no value in doing so in
  true peer review.
\item Some have pointed out the incremental, often poor quality of
  work that the publish or perish paradigm induces.  As a solution,
  their proposal is to make people understand that science happens
  slowly~\cite{slow-science:2010}.  I don't think the problem is one
  of speed.  The fundamental problem is gatekeeping.  If that is
  fixed, the problem of incremental, poor quality work will disappear.
\item True peer review will promote a more open community of
  researchers, one where researchers discuss and explore rather than
  write quick and often unduly harsh or superficial reviews under time
  duress as currently happens.  
\item In the pursuit of publications, it is \emph{teaching} that has
  been getting the short shrift.  Junior appointments are worried
  about having too high a teaching load because that would get in the
  way of publishing.  At some institutions, faculty members are able
  to trade grant money for reduced teaching loads.  At others,
  teaching is delegated to postdoctoral students and sometimes
  teaching assistants.  Universities know the value of publications,
  so they help faculty members with reduced teaching loads.  If the
  publication count were to become meaningless, then teaching would
  once again rise to the prominence it deserves.

  Teaching is \emph{no less important} than research.  Researchers who
  consider it of secondary importance do so at their own peril.  In
  their students, they have a fresh readily available audience for
  their ideas---day in, day out, nine months a year.  Among their
  students will most likely be the people who will in the future take
  their ideas and research program forward.

\end{itemize}

\section{Conclusions}

What I have tried to show in this paper is that traditional peer
review has almost nothing going for it except tradition whereas true
peer review has no foreseeable flaws.  Traditional peer review
conflates the notion of peer review with the idea of gatekeeping.
Gatekeeping is not an empirically validated activity; it just happens
to be the traditional system.  For those who argue that anything that
replaces gatekeeping should be better, the onus is on them to first
show how well gatekeeping performs.

If we go by the responses to PRC's
survey~\cite{ware:peer-review:2008}, it seems that a majority of
researchers find merit in peer review.  Among the benefits cited are
include improved quality, detection of fraud, and so on.  Each of the
cited benefits is potentially attainable to a greater degree in true
peer review but without all the hassles and pretensions that are
associated with the former.

Getting rid of gatekeeping means giving more importance to content,
which should have been our guiding value, but was instead lost in the
clamor for more and more publications.  It is disheartening when
researchers would rather work on finishing and submitting an
unpromising paper given an impending deadline rather than think and
talk about the underlying challenges.  It is disheartening that our
doctoral students are so burdened with producing papers that they
would rather not make any presentations in weekly seminars.  It is
disheartening to see them stumbling about in the dark, doing this and
that, but never really striving to get to the crux of the matter.
It's not their fault: one doesn't have to get to the crux of things to
get published.  It is disheartening to see that researchers are
actually turned off by spirited but definitely polite debate.  It is
disheartening that we have set up a system which discourages the
pursuit of knowledge.

If we get rid of gatekeeping, we will have to build \emph{everything}
anew.  Because right now, everything is built upon the idea that
gatekeeping has merit in that only the papers it selects have merit.
Getting rid of gatekeeping means substantial changes in
\emph{evaluations}.  It would affect the way we hire and promote
researchers, allocate funding, and award honors.  Here I would like to
emphasize one thing that I already addressed in some detail in
Section~\ref{sec:fa-gk}.  One argument people bring up again and again
is that true peer review is a pipe dream unless I can also show that
systems of evaluations will also work better in a true peer review
world.  It would be good if I had those answers but the legitimacy of
true peer review does not rest upon answering those questions.
Conceptually, I see that any system of evaluations would be
\emph{built upon} an underlying system whose value is the pursuit of
knowledge.  The first thing is to make sure that the underlying system
works in and of itself because if that system fails, then as computer
scientists know well enough, everything built on top will fail.  I
have argued that current peer review fails as this underlying layer
whereas true peer review succeeds.

Perhaps in the Internet-less age, gatekeeping served a purpose given
the practical limits on dissemination.  Now it hampers dissemination.
It shackles researchers and science.  Gatekeeping is nothing but an
exercise in futility, vanity, and censorship.  Let's get rid of it.

I think of each mind as a rich world of its own.  And I like to think
of true peer review as exploring a problem not only with your own mind
but also through the minds of others.  Knowledge is not out there; it
is hidden deep inside the pathways of our minds.  We can begin to get
to it only by talking, discussing, and reflecting.  The richness of
the ideas that would be born from exploring many minds would be truly
breathtaking.

\textit{Acknowledgments}: This work was supported by a Marie Curie
Trentino Cofund award and by the ERC Advanced Grant 267856
``Lucretius: Foundations for Software Evolution''.  I am grateful to
my colleagues in the Requirements Engineering Seminar Group at the
University of Trento, especially Fabiano Dalpiaz, Vitor Silva Souza,
Mohamad Gharib, and Jennifer Horkoff for extensive discussions.  Julio
Leite, Magda Altman, Mateus Joffily, Sinan Mutlu, Marta Facchini,
Michael Huhns, Stephen Cranefield, and Fausto Giunchiglia gave
extensive comments on earlier drafts.  Above all I am indebted to
Munindar Singh, from whom I have imbibed the importance of values.

\end{document}